\documentclass[aps,twocolumn,pra,showpacs]{revtex4}
\usepackage[T1]{fontenc}
\usepackage[dvips]{graphicx}
\usepackage[latin1]{inputenc}
\begin{document}
\begin{quotation}
Proceedings ICSSUR2001 
\end{quotation}
\vglue0.3in

\title{Is ``entanglement'' always entangled?}
\author{A. F. Kracklauer}
\email{kracklau@fossi.uni-weimar.de}

\begin{abstract}
Entanglement, including ``quantum entanglement,'' is a
consequence of correlation between objects. When the objects are
subunits of pairs which in turn are members of an ensemble
described by a wave function, a correlation among the subunits
induces the mysterious properties of ``cat-states.'' However,
correlation between subsystems can be present from purely
non-quantum sources, thereby entailing no unfathomable behavior.
Such entanglement arises whenever the so-called ``qubit space''
is not afflicted with Heisenberg Uncertainty. It turns out that
all optical experimental realizations of EPR's \emph{Gedanken}
experiment in fact do not suffer Heisenberg Uncertainty. Examples
will be analyzed and non-quantum models for some of these
described. The consequences for experiments that were to test
EPR's contention in the form of Bell's Theorem are drawn:
\emph{valid tests of EPR's  hypothesis have yet to be done.}
\end{abstract}


\maketitle

\section{Introduction}

The above title needs `disentanglement.' The quantum wave function of
entangled, i.e., of  correlated subsystems, can not be written as the product
of the wave functions for the subsystems. Likewise the probability of
correlated events can not be written as the product of probabilities for
two independent events. The latter fact is elementary and very well
understood; it presents absolutely no mystery, but in Quantum Mechanics (QM),
on the contrary, the same fact is utterly impenetrable. 

What is the difference? 

It arises from the following considerations. In probability theory, the probability
for joint events is given in general by Bayes' formula:
\begin{equation}
\label{Bayes}
P(a,\: b)=P(a)P(b|\: a),
\end{equation}
where \( P(b|\: a) \) is the conditional probability that the event \( b \)
occurs given that event \( a \) has been seen.\cite{wf} This is a statement
about the \emph{knowledge} that the observer has about the \emph{joint} events;
it is an \emph{epistemological} statement, and, as such, the dependance of
\(P(b|\: a) \) on \(a\) is devoid of communicative implications.  Now, in QM,
according to the Born interpretation, the modulus squared of a wave function,
i.e. \( \psi ^{*}(x)\psi (x) \), is the probability that the object to which it
pertains will be found in the infinitesimal volume \( d^{3}x \). This
straightforward concept is complicated, however, by \emph{the} peculiarity
of QM, namely, a wave function is known empirically to diffract at boundaries,
just like water or electromagnetic waves, and this seems to make sense only if
wave functions have \emph{ontological} substance. In turn, this appears to vest
a causative relationship into conditional probabilities computed from wave
functions for correlated events. In short, \emph{entanglement\( _{QM} \)} is
somehow ontological, but \emph{entanglement\( _{Prob} \),} epistemological. In
this light the title is: \emph{Is (in the microscopic domain) entanglement\(
_{Prob} \) always entangled\( _{QM} \)?} The purpose of this report is to argue
that in virtually all of the crucial experimental tests of Bell's Theorem, the
answer is: no!

Born's interpretation of the wave function has led many, in particular
Einstein, Podolsky and Rosen (EPR), to argue that the necessity for
probabilistic concepts in QM arises because the theory is limited fundamentally
by ignorance; i.e., that QM should be `extendable,' at least in principle, so
as to encompass the heretofore missing information, perhaps using ``hidden
variables.'' The tactic taken by EPR was to show that `Heisenberg Uncertainty'
is not something novel, that is, that basic logic regarding correlated objects
demands that the missing information be due to simple ignorance. This they did
by considering the symmetrical disintegration of a stationary particle into
twin daughters. For each daughter separately, the Heisenberg Uncertainty
Principle implies that both the position and momentum can not be simultaneously
known to arbitrary precision. Some go on to argue that this is so because they
in fact do not exist simultaneously. EPR countered, arguing (\emph{in the
author's rendition}) that in the case of such a disintegration one can measure
the position of one daughter and the momentum of the other to arbitrary
precision and thereafter call on symmetry to specify to equal precision the
momentum of the first and position of the second. What can be specified in
principle to arbitrary precision, EPR argued, must be an ``element of reality''
that enjoys ontological status. In any case, EPR intended that their
\emph{Gedanken} experiment should expose the ultimate character of Heisenberg
Uncertainty, that it is ultimately just ignorance, not something fundamentally
new.\cite{epr}

For the purposes of an experimental realization of EPR's \emph{Gedanken}
experiment, however, the difficulties finding a suitable source of the sort
envisioned, are daunting. Thus, Bohm proposed a change of venue; instead of
momentum-position, he suggested using the (anti)correlated spin states derived
from a mother with no net angular momentum.\cite{db} His motivation, apparently,
was that it should be easier to construct an appropriate source, and easier to
measure the dichotomic values of the daughters. Ultimately, this proposal too
turned out to be impractical, but the algebraically isomorphic situation with
polarized `photons' from a cascade transition or from parametric down
conversion is workable and a several such experiments have been
done.\cite{as}

\section{Entanglement\protect\( _{QM}\protect \) vs. entanglement\protect\( _{Prob}\protect \)}

It is the fundamental premise of this report that Bohm's transfer of venue
introduced a major error. It is the following: the space of the variables for
either spin or polarization, contrary to phase space where EPR formulated their
\emph{Gedanken} experiment, is not afflicted by Heisenberg Uncertainty (HU).
There is no HU in the plane of the spin or polarization vector. Neither \(
\{E_{x},\: E_{y}\} \) nor \{\( \sigma _{x},\: \sigma _{y}\} \) are Hamiltonian
canonically conjugate variables; their creation and annihilation operators
commute. Anticommutation of spin operators here arises for the same reason it
does for angular momentum operators in classical mechanics.  Thus, while they
do share some of the characteristics of the variables of phase space, they do
not share the one relevant for the argument of EPR.

This fact has a number of immediate consequences, the most salient of which is
that probabilities of these variables do not exhibit the quantum phenomena that
ultimately demands that QM probabilities have an ontological character. This
means, in particular, that conditional probabilities of these variables do not
imply causality. Thus, Bell's argument that because there is to be no
\emph{causal} relationship between the two detection events, the probability
relationship between them can not take the form \begin{equation} \label{hv}
P(a,\: b)=\int P(a|\: b,\: \lambda ))P(b|\: \lambda )P(\lambda )d\lambda ,
\end{equation} which, in turn, implies that Eq. (\ref{Bayes}) must read \(
P(a,\: b|\:\lambda)=P(a|\:\lambda)P(b|\:\lambda) \) \cite{jb}, does not follow
for these experiments, because, in fact there need be no causative link between
these variables.\cite{as} In other words, Bell's encoding of ``locality'' with
respect to these variables is not justified in these circumstances. A
conditional probability involving a state of polarization as a `condition' is
an epistemological statement about the state of knowledge, not an ontological
statement about EPR's ``elements of reality.'' This follows directly from the
fact that there is no reason whatsoever to attribute physical interference
between polarization states, these states are simply orthogonal from the start
and do not interact. In short, statements about joint probabilities between
such states do not imply any causal relationships; the non-factorizability of
their wave function is no more problematic than that of probabilities of
correlated events.

\section{Non-quantum models of EPR-B experiments}

In view of the facts developed above, which imply that experiments exploiting
polarization that are intended to test EPR (or Bell inequalities), in so far as
they are not cast in a space suffering HU, should be modelable classically.
This is indeed the case, and the most common types of EPR-B experiments are
presented below. These include both those based on polarization and a second
category in which orthogonality of the signals is achieved by other means,
usually as pulses with a phase offset. This latter category includes the
`Franson,' `Ghosh and Mandel' and `Suarez-Gisin' type experiments.

\subsection{`Clauser-Aspect' type experiments}

In these experiments the source is a vapor, typically of mercury or calcium, in
which a cascade transition is excited by either an electron beam or an intense
radiation beam of fixed orientation. Each stage of the cascade results in
emission of radiation (a ``photon'') that is polarized orthogonally to that of
the other stage. In so far as the sum of the emissions can carry off no net
angular momentum, the separate emissions are antisymmetric in space. The
intensity of the emission is maintained sufficiently low so that at any instant
the likelihood is that radiation  from only one atom is visible. Photodetectors
are placed at opposite sides of the source, each behind a polarizer with a
given setting. The experiment consists of measuring the coincidence count rate
as a function of the polarizer settings.\cite{ch}

A model consists of simply rendering the source and polarizers mathematically,
and a computation of the coincidence rate. Photodetectors are assumed to
convert continuous radiation into an electron current at random times with
Poisson distribution but in proportion to the intensity of the radiation. The
coincidence count rate is taken to be proportional to the fourth order
coherence function evaluated at the detectors. 

The source is assumed to emit a double signal for which individual signal components
are anticorrelated and, because of the fixed orientation of the excitation source,
confined to the vertical and horizontal polarization modes; i.e.
\begin{equation}
\label{30}
\begin{array}{cc}
S_{1} & =(cos(n\frac{\pi }{2}),\: sin(n\frac{\pi }{2}))\\
S_{2} & =(sin(n\frac{\pi }{2}),\: -cos(n\frac{\pi }{2}))
\end{array},
\end{equation}
 where \( n \) takes on the values \( 0 \) and \( 1 \) with an even, random
distribution. The transition matrix for a polarizer is given by,

\begin{equation}
\label{40}
P(\theta )=\left[ \begin{array}{cc}
\cos ^{2}(\theta ) & \cos (\theta )\sin (\theta )\\
\sin (\theta )\cos (\theta ) & \sin ^{2}(\theta )
\end{array}\right] ,
\end{equation}
 so the fields entering the photodetectors are given by:
\begin{equation}
\label{50}
\begin{array}{cc}
E_{1} & =P(\theta _{1})S_{1}\\
E_{2} & =P(\theta _{2})S_{2}
\end{array}.
\end{equation}
 Coincidence detections among \( N \) photodetectors (here \( N=2 \)) are
proportional to the single time, multiple location second order cross correlation,
i.e.:
\begin{equation}
\label{e2}
P(\theta_{1},\, \theta_{2},..\theta_{N})=\frac{<\prod ^{N}_{n=1}E^{*}(r_{n},\!
\theta)\prod ^{1}_{n=N}E(r_{n},\! \theta)>}{\prod
^{N}_{n=1}<E^{*}_{n}(r_n)E_{n}(r_n)>}.
\end{equation}
 It is shown in Coherence theory that the numerator of Eq. (\ref{e2}) reduces
to the trace of \( \mathbf{J} \), the system coherence or ``polarization''
tensor.\cite{mw} It is easy to show that for this model the denominator consists
of constants and will be ignored as we are interested only in relative intensities.
The final result of the above is:
\begin{equation}
\label{60}
P(\theta _{1},\theta _{2})=\frac{1}{2}\sin ^{2}(\theta _{1}-\theta _{2}).
\end{equation}
 This is immediately recognized as the so-called `quantum' result. (Of course,
it is also Malus' Law, thereby being in total accord with the premise of this
report.)

\subsection{`GHZ' experiments}

A number of proposed experiments involving more that two particles, many
stimulated by analysis of Greenburger, Horne and Zeilinger (GHZ), are expected
to reveal QM features with particularly alacrity.\cite{ghz} One of the most
recent, which has the great virtue of being experimentally doable, is that
performed by Pan et al.\cite{pd} See Fig. (1).  Two independent signal pairs
are created by down-conversion in a crystal pumped by a pulsed laser. The laser
pulse passes through the crystal creating one pair then is reflected off a
movable mirror to repass through the crystal in the opposite direction creating
a second pair. One signal from each pair is fed directly through polarizers to
photodetectors (signals \(A_1\) and \(B_1\). The other signal from each pair
\(A_2\) and \(B_2\) is directed to opposite faces of a PBS, (i.e., a beam
splitter which reflects vertically and transmits horizontally polarized
signals) after which the signals are passed through adjustable polarizers into
photodetectors. The path lengths of signals 2 and 3 are adjusted so as to
compensate for the time delay in the creation of the pairs. By moving the
mirror, the compensation can be negated to permit studying the coincidence
dependence on the degree of interference caused by simultaneous ``cross-talk''
between channels 2 and 3.

\begin{figure}[h]
\begin{center}
\includegraphics[width=0.70\columnwidth]{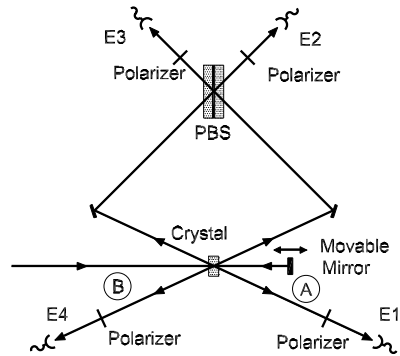}%
\end{center}
\caption{Schematic of the experimental setup for the measurement
of four-photon GHZ correlations. A pulse of laser light passes a nonlinear
crystal twice to produce two entangled photon pairs via 
parametric down conversion. Coincidences
between all four detectors are used to study the nature of 
quantum entanglement. }
\label{setup}
\end{figure}

The principle results reported in Ref. \cite{pd} are the following. Of all the
16 possible regimes setting: \( \theta _{i}=\{0,\: \pi /2\} \), only \( \{0,\:
\pi /2,\: \pi /2,\: 0\} \) and \( \{\pi /2,\: 0,\: 0,\: \pi /2\} \) yield a
(substantial) four-fold coincidence count, \( C \); the regime \( \{\pi /4,\:
\pi /4,\: \pi /4,\: \pi /4\} \) occurs with an intensity \( C/4 \) and the
regime \( \{\pi /4,\: \pi /4,\: \pi /4,\: -\pi /4\} \) with zero intensity.
Further, both of the later regimes yield an intensity of \( C/8 \) when the
time between pair creation is so large that that there is no ``cross-talk''
between channels 2 and 3.

\begin{figure}[b]
\begin{center}
\includegraphics[width=0.99\columnwidth]{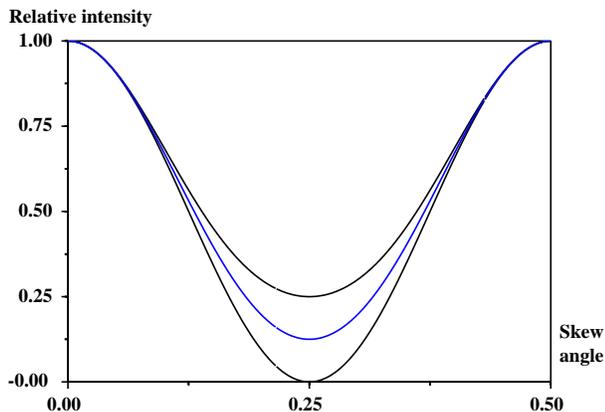}  
\end{center}
\caption{The upper curve shows the effect on the intensity of four-fold 
coincidences of skewing (rotating) all polarizers through a given angle in
units of $\pi$-radians starting from the state $\{\pi/2,0,0,\pi/2\}$.   The
lower  curve shows the same effect when one of the polarizers is rotated in 
the opposite direction. The middle curve shows the effect of either of  these
skewing schemes when the timing is such that the  crossover signals do not
arrive simultaneously with the reflected signals. Note that the values at
\(\pi/4 \) coincide with those observed.  This diagram differs from
Fig. 4 in Ref. \cite{pd} in that it shows the split of these regimes as a
function of polarizer skew for fixed delay rather than  as a function of delay
for fixed skew.}
\label{skew}
\end{figure}

Eq. (\ref{e2}) was implemented as follows: The crystal is assumed to emit a
double signal for which individual signal components are anticorrelated and
confined to the vertical and horizontal polarization modes; i.e.
\begin{equation}
\label{70}
\begin{array}{cc}
A_{1}= & (cos(n\frac{\pi }{2}),\: sin(n\frac{\pi }{2}))\\
A_{2}= & (sin(n\frac{\pi }{2}),\: -cos(n\frac{\pi }{2}))\\
B_{1}= & (cos(m\frac{\pi }{2}),\: sin(m\frac{\pi }{2}))\\
B_{2}= & (sin(m\frac{\pi }{2}),\: -cos(m\frac{\pi }{2}))
\end{array}
\end{equation}
where \( n \) and \( m \) take the values \( 0 \) and \( 1 \) randomly.
The polarizing beam splitter (PBS) is modeled using the transition matrix for
a polarizer, \( P(\theta ) \), Eq. (\ref{40}) where \( \theta =\pi /2 \)
accounts for a reflection and \( \theta =0 \) a transmission. Thus the final
field impinging on each of the four detectors is :
\begin{equation}
\label{80}
\begin{array}{ll}
E_{1}= & P(\theta _{1})A_{1}\\
E_{2}= & P(\theta _{2})(P(0)B_{2}-P(\pi /2)A_{2})\\
E_{3}= & P(\theta _{3})(P(0)A_{2}-P(\pi /2)B_{2})\\
E_{4}= & P(\theta _{4})B_{2}
\end{array}
\end{equation}
which, using Eq. (\ref{e2}), does not result in a simple expression. However,
it can be numerically computed easily to obtain the same results as reported
by Pan et al., or extended to other regimes, such as that shown in Fig. (2).

\subsection{`Franson' experiments}

Experiments of this type exploit phase shifts between pulses in the form of
time offsets to define the orthogonal states played by the two states of
polarization in the setups described above.\cite{jf} The original `Franson'
experiment measures the correlation between two detectors positioned after
interferometers which divide identical incoming pulses such that half takes a
short route and half takes a long root which includes an adjustable delay. See
Fig. (3). 

\begin{figure}[h]
\begin{center}
\includegraphics[width=0.99\columnwidth]{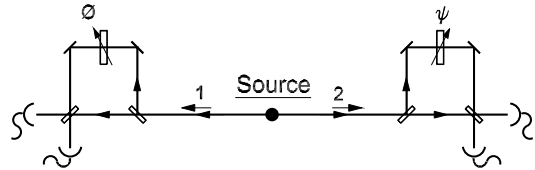}  
\end{center}
\caption{In a `Franson' type experiment two identical pulses are directed
through two interferometers, each comprised of a short path and a long path in
which there is an additional adjustable phase shifter. By using fast
coincidence comparison detectors, coincidences between pulses that traversed
unequal paths can be excluded. The resulting interference is a function
of the adjustable  phase shifters.}
\label{bild}
\end{figure}

There are two means of modeling this setup. One would be to write out terms for
the long- and short-route pulses that had time-separated modulation or
time-limited coherence. This approach has the disadvantage of leading ungainly
expressions. A much simpler tactic is to assign the signals in the long and
short paths to orthogonal dimensions of a vector space; the resulting
calculations are then transparent and devoid of gratuitous
complexity. For example:

\begin{equation}
\label{90}
\begin{array}{cc}
E_{l} & =(\exp (-i(kx-\omega t)+\phi ),\: \exp (-i(kx-\omega t))/\sqrt{2}\\
E_{r} & =(\exp (-i(kx-\omega t)+\psi ),\: \exp (-i(kx-\omega t))/\sqrt{2}
\end{array},
\end{equation}
 where \( \phi  \) and \( \psi  \) are the extra phase shifts introduced
in the long paths. Then, using Eq. (\ref{e2}), with the convention that the
tensor product be replaced by a vector inner product; i.e.,

\begin{equation}
\label{91}
P(\phi ,\: \psi )=\frac{(E^{*}_{r}\cdot E^{*}_{l})(E_{l}\cdot E_{r})}{(E^{*}_{r}\cdot E_{r})(E^{*}_{l}\cdot E_{l})},
\end{equation}
 (to algebraically enforce the orthogonality in calculations that phase shifts
enforce in the experiment) directly gives the observed correlation as a function
of the phase shifts: 
\begin{equation}
\label{100}
P(\phi ,\: \psi )\propto(1+\cos(\phi -\psi )),
\end{equation}
 which exhibits the oscillation with \( 100\% \) visibility characteristic
of idealized versions of these experiments. 

`Ghosh-Mandel' type experiments are a variation of the `Franson' version in
which the phase shift is achieved by path-length differences instead
of time-offsets; otherwise, the formulas are identical.\cite{GM}

\subsection{`Brendel' experiments}

In the above experiment the radiation source was taken to be ideal, that is, it
produced two signals of exactly the same frequency with no dispersion. In some
experiments, \cite{bm} the source used was a nonlinear crystal generating two
correlated but not necessarily identical pulses, which satisfy `phase matching
conditions' so that if one signal in frequency is above the mean by \( s \)
(spread), the other is down in frequency by the same amount. This leads to an
additional phase shift at the detectors which is also proportional to
those already there; i.e., \( s\phi \) and \(- s\psi  \), so that: 

\begin{equation}
\label{Brendel}
\begin{array}{cc}
E_{r} & =(\exp (-i(kx-\omega t)+\psi (1+s)),\: \exp (-i(kx-\omega t))\\
E_{l} & =(\exp (-i(kx-\omega t)+\phi (1-s)),\: \exp (-i(kx-\omega t))
\end{array}.
\end{equation}
 Since the value of \( s \) is different for each pulse (photon) pair, the
resulting signal is an average over the relevant values of \( s \):
\begin{equation}
\label{BR1}
\frac{1}{2s}\int ^{s}_{-s}P(\phi ,\: \psi ,\: s)ds,
\end{equation}
 where \( P(\phi ,\: \varphi ,\: s) \) is computed as for `Franson' experiments.
The final result closely matches that observed by Brendel et al. See Fig. (4).

\begin{figure}[h]
\begin{center}
\includegraphics[width=1.2\columnwidth]{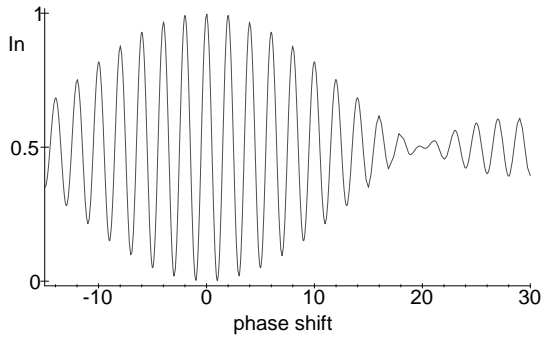}  
\end{center}
\caption{Plot of the relative interference intensity (In) pattern as a function
of phase shift (in units of \(\pi\)) in one arm in a Brendel type
experiment.  This curve closely matches that observed by Brendel et al. in an
experiment in which the total spread was \( 10\% \) of the pulse carrier
frequency; as a result, the modulation curve node occurs at approximately \(20
\pi \), as was observed.}

\label{BRE}
\end{figure}

\subsection{`Suarez-Gisin' experiments}

In experiments of this type, one
of the detectors is set in motion relative to the other. By doing so with
appropriately chosen parameters, it is possible to arrange the situation such
that each detector precedes the other in its own frame.\cite{gs} Thus, not only
is the `collapse' of the wave packet ``nonlocal,'' it occurs such that there is
also ``retrocausality.'' In the model proposed herein, however, this
complication (paradox) can not arise in the first instance. All the properties
of each pulse are determined completely at the common point at which the
signals are generated. Properties measured at one detector in no way determine
those at other detectors, regardless of the order in which an observer receives
reports of the results from various detectors, and regardless of what
conditional probabilities he might write to describe his hypothetical or real
knowledge.

\section{Conclusion}

The model or explanation of the experiments described above is fully classical.
It uses no special property peculiar to QM. The two states in these experiments
(polarization or phase-displaced pulses) are not canonically conjugate dynamical
variables; they do not, therefore, exhibit Heisenberg Uncertainty, and the
model does not bring any in. The essential formulas are a straightforward
application of second order (in intensity) coherence theory, which is really
just a generalization of wave interference. That this model faithfully
describes the outcomes of these experiments, in addition to being a
counterexample to claims that these experiments can not be clarified using
non-quantum physics, is a demonstration that they are not relevant to EPR's
argumentation, and therefore, that to date no such experiment could have
established that non-locality has a role to play in the explanation of the
natural world. It shows that there is no justification for ascribing an
ontological meaning to conditional probabilities in the circumstances of these
experiments, which, in turn, undermines the rationale for Bell's encoding of
non-locality. When his encoding is withdrawn, no Bell inequality can be
extracted. 

There are, of course, two arenas where HU is in evidence: phase space and `quadrature
space.' In principle, a test of EPR's contentions formulated in these arenas
could show different results --- at least in so far as the considerations herein
are germane. 

To a large extent, the model proposed herein is `obvious.' It might be asked:
why has it not been proposed then long ago? The answer involves issues resulting
from the perceived need to maintain an ontological ambiguity with respect to
the identity of wave functions until the moment of measurement, at which time
this identity ambiguity is resolved by a ``collapse.'' This need results from
the tactic of describing \emph{particle} beams with wave functions in order
to account for their wave-like diffraction. That is, the wave-like navigation
of particle beams in combination with their incontestable particle-like registration
in detectors, has been explained, or at least encoded, calling on `dualism,'
`wave-collapse' and so on.\cite{ak} The experiments described herein, however,
employ optical phenomena for which there is no need to invoke a particulate
\emph{}character. Wave beams diffract naturally. And, particulateness in detectors
can be, indeed must be, attributed to the fact that photodetectors, because
of the discrete nature of electrons, convert continuous radiation into a digitized
photocurrent. The conceptual contraptions of `duality' and `collapse' are just
not needed to explain the behavior of radiation beams, even correlated sub beams.
There is no reason these experiments could not be carried out in spectral regions
in which it is possible to track the time development of electromagnetic fields
thereby avoiding the peculiarities of photodetectors. In fact, for simple `Clauser-Aspect'
type setups, this has been done.\cite{ek} The results conform with ours and
show that classical optics is not taxed to clarify EPR-B correlations.

\emph{Note:} An e-file with MAPLE routines for the above is available upon
request.

\end{document}